\documentclass[12pt,preprint]{aastex}
\usepackage{natbib}

\shorttitle{A MSP in a very eccentric binary system}
\shortauthors{Freire et al.}

\begin{document}

\title{GMRT Discovery of A Millisecond Pulsar in a Very Eccentric
  Binary System}

\author{Paulo C. Freire\altaffilmark{1},
  Yashwant Gupta\altaffilmark{2},
  Scott M. Ransom\altaffilmark{3,4},
  Ishwara-Chandra C. H.\altaffilmark{2}}
\altaffiltext{1}{NAIC, Arecibo Observatory, HC3 Box 53995, PR 00612,
  USA; {\tt pfreire@naic.edu}}
\altaffiltext{2}{National Centre for Radio Astrophysics, P.O. Bag 3,
  Ganeshkhind, Pune 411007, India; {\tt ygupta@ncra.tifr.res.in,
  ishwar@ncra.tifr.res.in}}
\altaffiltext{3}{Department of Physics, Rutherford Physics Building,
  McGill University, 3600 University Street, Montreal, Quebec, H3A
  2T8, Canada; {\tt ransom@physics.mcgill.ca}}
\altaffiltext{4}{Department of Physics and Center for Space Research,
  Massachusetts Institute of Technology, Cambridge, MA 02139}


\begin{abstract}
  We report the discovery of the binary millisecond pulsar
  J0514$-$4002A, which is the first known pulsar in the globular
  cluster NGC~1851 and the first pulsar discovered using the Giant
  Metrewave Radio Telescope (GMRT). The pulsar has a rotational period
  of 4.99\,ms, an orbital period of 18.8\,days, and the most eccentric
  pulsar orbit yet measured ($e = 0.89$). The companion has a minimum
  mass of 0.9\,M$_{\odot}$ and its nature is presently unclear. After
  accreting matter from a low-mass companion star which spun it up to
  a (few) millisecond spin period, the pulsar eventually exchanged the
  low-mass star for its more massive present companion.  This is
  exactly the same process that could form a system containing a
  millisecond pulsar and a black hole; the discovery of NGC~1851A
  demonstrates that such systems might exist in the Universe, provided
  that stellar mass black holes exist in globular clusters.
\end{abstract}


\section{Introduction}\label{sec:intro}

Since 1987, several globular cluster surveys at Jodrell Bank, Arecibo,
Parkes and more recently the Green Bank telescope have found 79
pulsars in a total of 23 globular clusters (GCs)\footnote{See
  http://www2.naic.edu/$\sim$pfreire/GCpsr.html and references
  therein}.  These findings have confirmed that most of the binary
millisecond pulsars (MSPs) in GCs have low-mass white dwarf (WD)
companions and nearly circular orbits, as observed in the Galactic
disk. This is an important confirmation of the evolutionary scenarios
proposed by Alpar~et~al.~(1982)\nocite{acrs82} for the formation of
MSPs.
 
In GCs, {\em exchange encounters}, which only have a significant
probability of occurring in dense stellar environments, occasionally
exchange one of the components of a binary system with a typically
more massive star.  The exchanges may occur during encounters with
either other binaries or with isolated stars.  In GCs such encounters
can place isolated neutron stars into binaries with a main sequence
(MS) star which eventually evolves, ``recycles'' the neutron star, and
finally forms a MSP$-$WD binary system.  Such a process explains the
anomalously large numbers of MSPs in GCs (by mass) when compared to
the Galaxy. If the neutron star placed in orbit around a low-mass MS
star is already a pulsar, we observe ``irregular'' eclipsing binary
pulsars, such as PSR~J0024$-$7204W in 47~Tucanae \cite{egc+02},
PSR~B1718$-$19 in NGC~6342 \cite{lbhb93}, and PSR~J1740$-$5340 in
NGC~6397 \cite{fpds01}, which are peculiar to GCs.

In this paper we report and discuss the discovery of a unique binary
millisecond pulsar, PSR~J0514$-$4002A (henceforth NGC~1851A) in the GC
NGC~1851. We discovered this system in a new 327-MHz survey of GCs
carried out using the Giant Metrewave Radio Telescope (GMRT), at
Khodad near Pune, India.

\section{The GMRT survey observations and the discovery of NGC1851A}
\label{sec:discovery}

Our 327-MHz survey aims to find faint pulsars with steep radio spectra
which would be missed by the high-frequency searches. It benefits from
the large gain of the central array of the GMRT of 4.6\,K/Jy when used
in the phased array mode of the Array Combiner
\cite{ggj+00,pra97}. This mode produces a
beam on the sky with a diameter of about 3 arcminutes.  The number of
spectral channels across the available 16-MHz band is 256 and the
sampling interval used is 258$\,\mu$s.  Each observation consists of a
pair of 72-minute scans containing 2$^{24}$ samples.  Between scans,
the 14-antenna central array was re-phased using a reference source.

We observed a set of 16 GCs in 2003 February. The data were written to
tape and taken to McGill University, where we processed them using the
BORG (a 104-processor Beowulf cluster dedicated for pulsar processing)
running the {\tt PRESTO} software package \cite{ran01}.  One of the
GCs observed was NGC~1851. Its distance ($D$) from the Sun is about
12.6~kpc (Cassisi, De Santis, \& Piersimoni 2001)\nocite{csp01}, and
its Galactic coordinates are $l = 244.51^\circ, b = -35.04^\circ$
\cite{har96}\footnote{See
  http://physwww.physics.mcmaster.ca/$\sim$harris/mwgc.dat for the
  updated version of the table of GC parameters presented in this
  paper}. It is a relatively bright GC ($M_v\,=\,-8.33$) with a very
condensed core ($c = log(r_t/r_c) = 2.32$, where $r_t$ and $r_c$ are
the tidal and core radii). It is among the ten clusters in the Galaxy
with the highest central luminosity density ($\rho_0\,\simeq\,2 \times
10^5\,\rm\,L_{\odot}\,pc^{-3}$).

Neither of the prominent electron density models of the
Galaxy\nocite{tc93,cl02} make reliable predictions of the electron
column density (normally expressed as the Dispersion Measure, or DM)
to NGC~1851. The Taylor \& Cordes (1993)
model predicts a DM$\sim$28\,pc\,cm$^{-3}$ for $D\,>\,3$\,kpc, while
the Cordes \& Lazio (2002; NE2001) model predicts
DM$\sim$45\,pc\,cm$^{-3}$ for $D\,>\,7$\,kpc. We therefore dedispersed
the raw data at 500 trial DMs,  from 20 to 70\,pc\,cm$^{-3}\,$ and
spaced by 0.1~cm$^{-3}\,$pc in order to minimize dispersive
smearing and keep the total time resolution of our data between 0.5
and 1.5\,ms.  For each of these DM trials, a full Fourier-domain
matched-filter-based acceleration search (Ransom, Eikenberry, \&
Middleditch 2002)\nocite{rem02} was carried out, enabling the
detection of binary systems with relatively short orbital periods.
This is only possible because of the computing power of the BORG.

\begin{figure}
\setlength{\unitlength}{1in}
 
\begin{picture}(0,7.0)
\put(1.3,8.0){\includegraphics{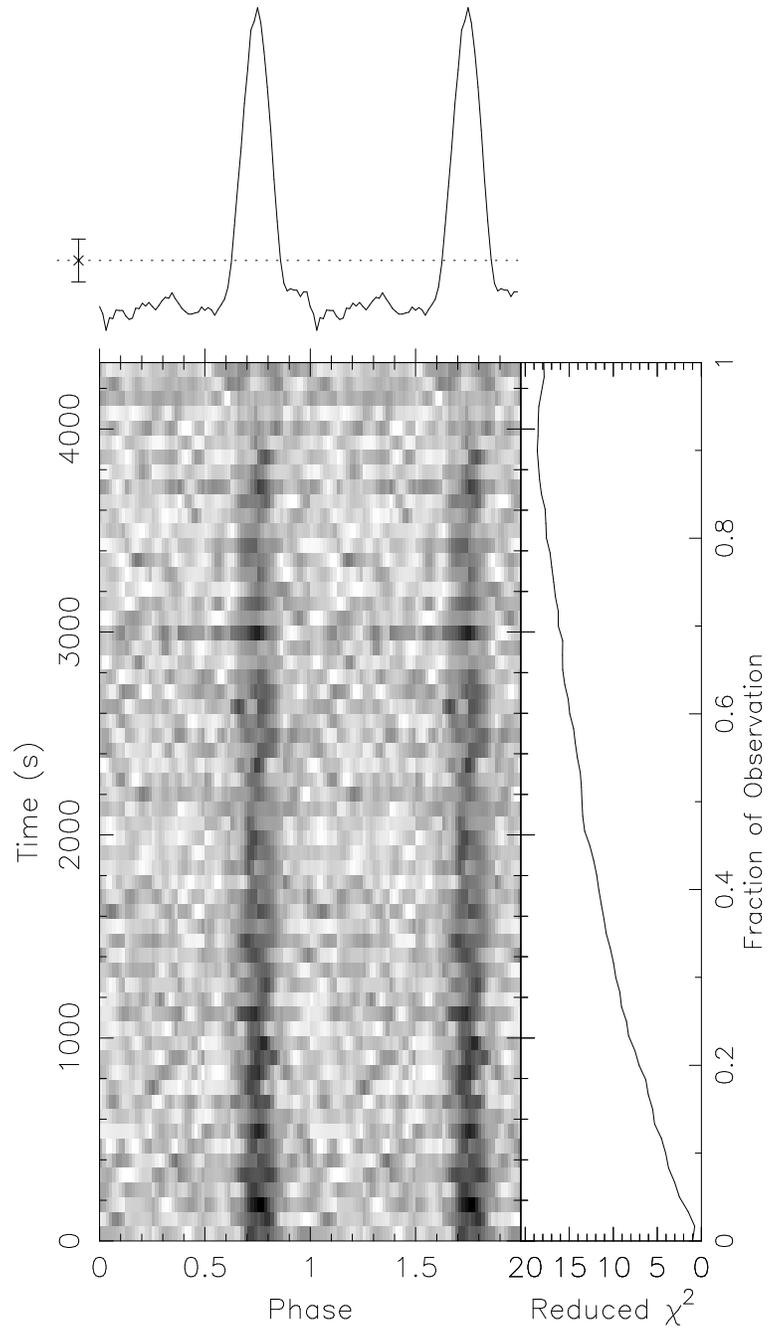}}\end{picture}
\caption [] {\label{fig:0514}
  Discovery observation for NGC~1851A, in the globular cluster
  NGC~1851. Pulsed emission is persistent throughout the scans. The
  pulse profile is rather narrow (top), corresponding to the time
  resolution of the system for this DM.}
\end{figure}

In the first scan for NGC~1851, taken 2003 February 10, we detected
a clear pulsed signal with a period of 4.991\,ms (see Figure
\ref{fig:0514}). The signal was present at all frequency channels
within the observing band and the arrival times versus frequency are
well described by a DM of 52.15(10)\,pc\,cm$^{-3}$. This value will
help to refine electron density models of the Galaxy in the direction
of NGC~1851. The observed pulsations had a width of $\sim$0.8 ms,
which is exactly the time resolution of the system after accounting
for the dispersive smearing within each of the 62.5-kHz channels and
the sampling interval. The intrinsic pulse is therefore essentially
unresolved, implying that future observations with better time and
frequency resolution will significantly improve the signal-to-noise
and timing precision.
We confirmed this signal in NGC~1851 data taken on the February 10, 11
and 20, the signal is not present in pointings made towards other
clusters. This is therefore the first pulsar discovered in NGC~1851 and,
incidentally, the first GMRT pulsar discovery. Highly
significant changes of the barycentric pulsed period both within and
between observations of the cluster immediately identified the pulsar
as a member of a binary system.
Further observations of NGC~1851 were made in 2003 December in
order to determine the orbit of the pulsar, using the same
observing system and parameters as for the original search.
An early analysis of the orbital trajectory of the pulsar in
the $P$-$\dot{P}$ diagram (Freire, Kramer, \& Lyne 2001)\nocite{fkl01}
suggested that the orbit is not circular.

\begin{figure*}
\setlength{\unitlength}{1in}
\begin{picture}(0,3.5)
\put(0.3,-0.3){\includegraphics{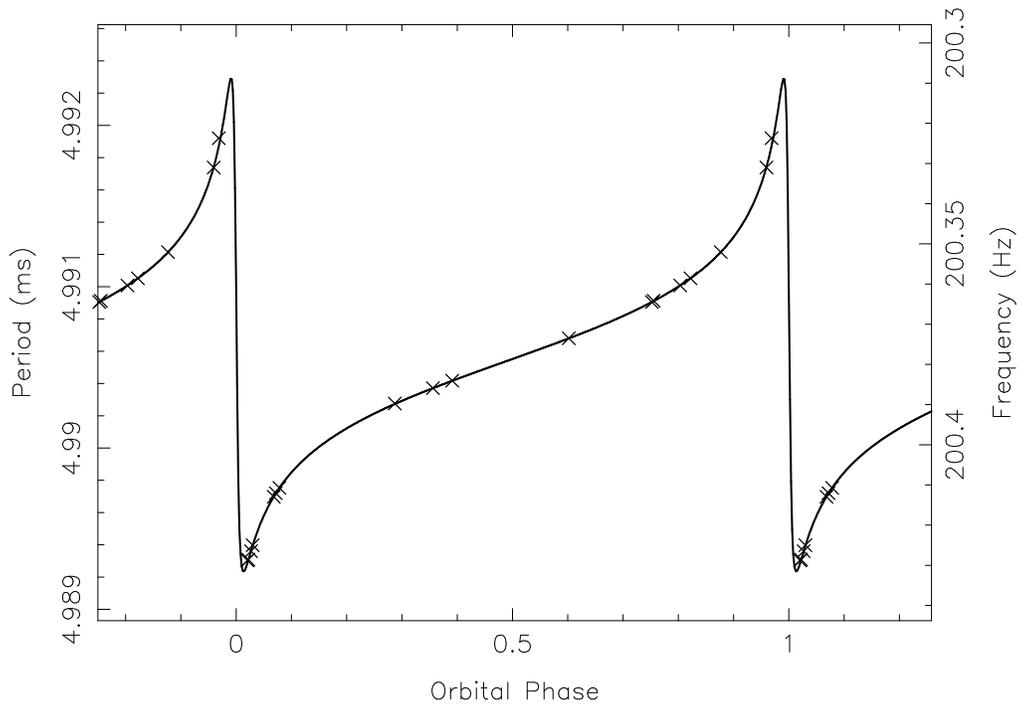}}
\end{picture}
\caption [] {\label{fig:periods}
  The measured rotational periods of NGC~1851A, as would be observed
  at the barycenter of the solar system, as a function of the pulsar's
  orbital phase. The solid lines indicate the prediction of the
  best-fit model.}
\end{figure*}

Analysis of the rotational periods using {\tt
  TEMPO}\footnote{\url{http://pulsar.princeton.edu/tempo}} proved most
surprising: the best-fit model (see Figure~\ref{fig:periods})
indicates $e\,=\,0.889(2)$, the most eccentric orbit of any known
binary pulsar, and several orders of magnitude more eccentric than the
typical orbits of fast MSPs\footnote{for this and following quantities, the
number in parenthesis indicates the 1-$\sigma$ uncertainty, which we
conservatively estimate to be ten times the formal value computed by
{\tt TEMPO} based on the fact that there is still no phase-connected
timing solution}. The orbital period $P_b$ is 18.7850(8) days, and the
semi-major axis of the orbit projected along the line-of-sight ($x$)
is $36.4(2)$ light seconds. This implies a minimum companion mass of
0.9\,M$_{\odot}$, assuming a pulsar mass of 1.35\,M$_{\odot}$ (for the
median of expected inclinations, 60$^\circ$, the companion mass is
1.1\,M$_{\odot}$). The epoch of periastron $T_0$ is MJD = 52984.46(2),
the longitude of periastron $\omega$ is $82(1)^\circ$ and the
barycentric rotational period $P$ is 4.990576(5) ms.

\section{Flux density and position}

\begin{figure*}
\setlength{\unitlength}{1in}
\begin{picture}(0,6)
\put(-1.2,+6.5){\includegraphics{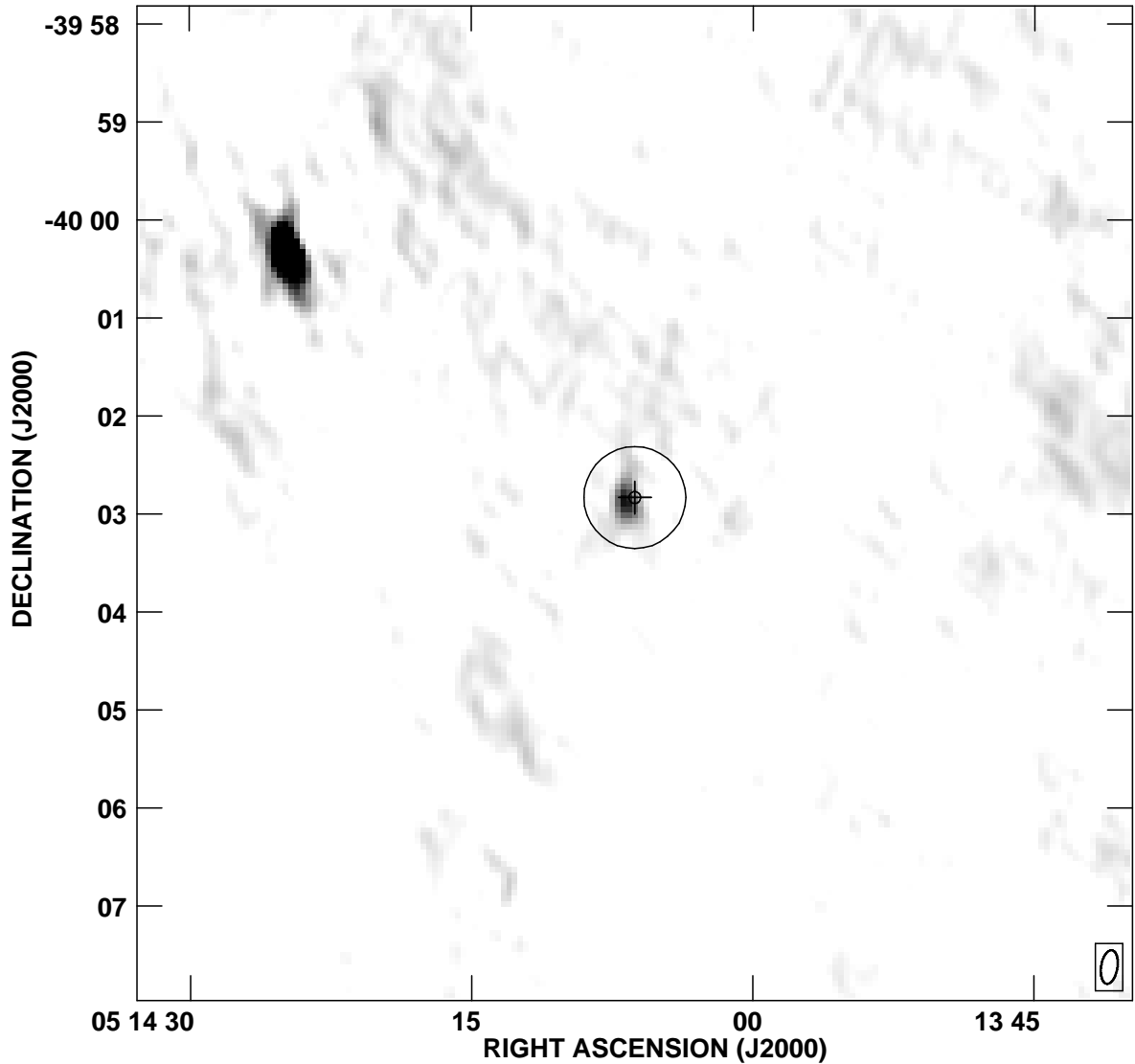}}
\end{picture}
\caption [] {\label{fig:image}
  A radio image of NGC~1851 at 327 MHz, made by combining the data
  from 4 different epochs of observations in December 2003. The pulsar
  is the faint source near the center of the cluster, which is
  indicated with a cross. The two circles indicate the core and
  half-mass radii (0.06 and 0.52 arcminutes respectively). The
  brighter source to the north-east is not the pulsar; when the
  interferometer was pointed at it no pulsed emission was detected.
  The small ellipse in the lower right corner indicates the dimensions
  of the synthesized beam of the interferometer, which includes {\em
    all} the active antennas. The diameter of the beam of the central
  square, used in the pulsar search, is about 3 arcminutes.}
\end{figure*}

When the pulsar is observed, the
cross-correlations between the voltages of all active antennae
(including those outside the central square, which are
not directly used to search for the pulsar) are also recorded.
This allows for simultaneous high-resolution imaging of the cluster.
We found a faint
object at $\alpha = 05^{\rm h}14^{\rm m}06.74\pm0.06^{\rm s}, \delta =
-40^\circ 02'50.0\pm1.3''$ (see Figure~\ref{fig:image}), which is only
0.1\,arcminutes from the center of the cluster ($\alpha = 05^{\rm
  h}14^{\rm m}06.2^{\rm s},\,\delta = -40^\circ 02'50''$, Harris
1996\nocite{har96}).  This places the source, as projected in the plane
of the sky, just outside the cluster's $\sim$0.06-arcminute core.
Such a position is common among the pulsars with known positions
within GCs as is expected due to the effects of
dynamical friction and mass segregation\cite{ka96,fcl+01}.

The flux density $S$ for this object is variable. For example, on the
5 December image, the source is not detectable (4$\sigma$ limit of
2.4\,mJy), while on 6, 12 and 30 December it is detectable with flux
densities of 5.5$\pm$0.5, 3.5$\pm$0.5 and 4.2$\pm$0.5\,mJy,
respectively.  This correlates well with the relative intensities of
the 4.99-ms pulsed signal for those days. The brighter source to the
north-east, on the other hand, has a constant flux density (within
10\%, which is the typical accuracy of the flux calibration scheme).
There is therefore little doubt that the faint source near the center
is NGC~1851A. The image in Figure \ref{fig:image} is a combination of
these four epochs and gives a final estimate of the mean flux density
at 325 MHz as 3.4$\pm$0.4\,mJy.  The image shows no other sources of
comparable flux density towards the cluster, indicating that we are
not missing the detection of bright pulsars because of selection
effects such as very short rotational or orbital periods.
We also observed the cluster at 610\,MHz on 7 and 23 December, as well
as 1 January. We did not detect the pulsar in any of the individual
epochs nor in a combined map made from all three observations at a
4-$\sigma$ level or better.  From the rms noise of 0.1\,mJy, we
estimate the 610-MHz flux density to be $<$0.4\,mJy.  This implies a
very steep spectrum, with $\alpha\,<\,-3.4$.
This underlines the relevance of low-frequency
surveys; it is unlikely that this pulsar could have been found at any
other frequency.

\section{Formation and nature}
\label{sec:formation}

All known eccentric ($e > 0.1$) binary pulsars in the disk of the
Galaxy have relatively massive companions. This varied set of
companions includes blue giants, other neutron stars and heavy WDs.
Blue giants live only a few Myr which is probably not long enough to
allow the sustained mass accretion required to spin up a pulsar
companion to millisecond spin periods.  This is in accordance with the
observations, these pulsars have
rotational periods of tens or hundreds of ms.
The second supernova event is likely to make the orbit significantly
eccentric, presuming the binary survives. Since both stars are now
compact, tidal circularization is henceforth impossible.  The
prolonged episode of stable mass accretion needed to spin up a neutron
star to millisecond periods is only possible from evolved lower-mass
MS stars. Such large timescales allow effective tidal orbit
circularization as well. This process likely created the MSP currently
in NGC~1851A, although with a presently unknown low-mass WD companion.

Some MSP $+$ low-mass white dwarf systems in GCs, like
PSR~B1802$-$07 ($P = 23.1$\,ms, $P_b = 2.62$\,days, $e = 0.212$,
\cite{dbl+93}) can become mildly eccentric due to interactions with
other objects in the cluster \cite{rh95}. This is almost certainly not
the origin of the present NGC~1851A binary system since its
eccentricity is probably too high to be explained by this mechanism
and its companion too massive. We are therefore lead to the conclusion
that after recycling, NGC~1851A exchanged its former low-mass
companion with a more massive object. This is the first system
presenting clear evidence of such a process. A massive star (the
pulsar's present companion) passed within a distance smaller than
about four times the separation of the components of the previous
binary system. The most likely outcome from such an event is the
formation of an eccentric binary system containing the two more
massive objects \cite{hut96b}. The low-mass component of the previous
binary is ejected, causing the new binary system to recoil in the
opposite direction. The probability for encounters is obviously
larger for wider binaries, but the resulting recoil velocities are
smaller.  These events lead to relatively wide binaries (as observed
for NGC~1851A) which are likely to remain near the center of the GCs.
This is consistent with the projected position of NGC~1851A relative
to the center of the cluster.  If the original binary system is
tightly bound, the resulting new binary system is also likely to be
very tight.  Such an encounter would produce a large recoil velocity
and could make the resulting system escape the cluster, or send it to
the cluster's outer regions.

There are three previously known eccentric pulsar binaries with
massive companions in GCs, but none of them presents such strong
evidence of this kind of exchange encounter.  The first is
PSR~B2127$+$11C \cite{agk+90}, a double neutron star in
M15 ($P\,=\,30.53\,$\,ms, $e\,=\,0.6813$, $P_b\,=\,0.33\,$\,days).
This system is remarkably similar to the first known Galactic binary
pulsar, PSR~B1913$+$16 \cite{wt03}; therefore no mechanisms specific
to GCs are needed to explain its formation. However, the fact that
this pulsar is much more distant from the center of M15 than any of
the other 7 pulsars known in that cluster \cite{and92} hints at an
exchange encounter with a powerful recoil from the ejected object.
The second system is PSR~J1750$-$37, \cite{pdm+01}. Its main
characteristics ($P\,=\,111.6\,$ms, $e\,=\,0.71$,
$P_b\,=\,17.3\,$days) are very similar to another Galactic binary
pulsar, PSR~J1811$-$1736 \cite{lcm+00}. An exchange
encounter is, again, not the only possible formation mechanism, but
definitely a possibility.  The third system is
PSR~J2140$-$2310B, in M30 \cite{rsb+04}, which contains a 13-ms pulsar
and $e\,>\,0.5$. Its rotational period is about half of the smallest
rotational period found for the eccentric Galactic systems. The
formation of such an object is very likely to require an exchange
encounter, but there is still a small probability that this is a MSP -
massive WD system that became eccentric through distant encounters
with other stars, like PSR~B1802$-$07.

The nature of the companion of NGC~1851A
is as yet unclear, it could be either a compact or extended
object. In NGC~1851, a cluster with an age of $\sim$9\,Gyr \cite{sw02}
1-M$_{\odot}$ stars are now leaving the main sequence. Because of the
lengthy episode of MSP recycling that preceded its formation, the
present NGC~1851A binary system is very likely to be a few Gyr younger
than the cluster in which it lies. Mathieu, Meibom and Dolan
(2004)\nocite{mmd04} have determined that, for the open cluster
NGC~188 (with an age of 7 Gyr and a stellar population similar to that
of GCs), binary systems containing MS stars with orbital periods
larger than 15 days have not yet had time to circularize. Therefore,
the observed eccentricity of the NGC~1851A system does not rule out
the possibility of the companion being an extended object.  In fact,
partial and/or irregular ``eclipses'' from an extended object such as
a MS star may explain the apparent flux variability from NGC~1851A.

\section{Conclusion}

We have discovered a remarkable 5-ms binary pulsar, the first to be
found in the GC NGC~1851 and the first pulsar to be discovered with
the GMRT. Its orbit is the most eccentric known for any system
containing a pulsar, while its rotational period is much shorter than
that of any other pulsar in an eccentric binary system.  This
indicates that, after becoming an MSP
by accreting matter from a low-mass companion star, this neutron star
has almost certainly exchanged it for its present, significantly more
massive companion. If black holes exist in GCs, a MSP-black hole
binary could be formed in exactly the same way.

If the companion is a compact object, then two relativistic effects
will be measurable, namely the rate of advance of periastron and the
Einstein delay. The measurement of both effects would lead to the
determination of the masses of the two components of this system. A
third relativistic effect, the orbital decay due to emission of
gravitational waves, will be masked by the unpredictable acceleration
of the binary in the gravitational field of the cluster, so no tests
of general relativity are likely to be possible. If the companion is a
MS star, we might measure the orbital precession caused by the quadrupole
moment of the companion star and possibly a rotation of the orbital
plane around the axis of rotation of the companion \cite{wjm+98}.  In
addition, variations in DM with orbital phase may be detectable
\cite{jml+96}, particularly if the companion is an active star. For
NGC~1851A none of these effects have been conclusively measured to date.

\acknowledgements

We thank J. Coppens, B. Bhattacharya and the staff of the GMRT for
help with the observations.  The Giant Metrewave Radio Telescope is
run by the National Center for Radio Astrophysics of the Tata
Institute of Fundamental research.  SMR acknowledges the support of a
Tomlinson Fellowship awarded by McGill University. The computing
facility used for this research was funded via a New Opportunities
Research Grant from the Canada Foundation for Innovation. We also
thank C. Salter and A. Deshpande for their suggestions
concerning the present paper.

\clearpage

\end{document}